\documentclass[runningheads]{llncs}
\usepackage{graphicx}
\usepackage{amsfonts}
\begin{document}
\title{Towards Semantic Interpretation of Thoracic Disease and COVID-19 Diagnosis Models}
\titlerunning{ }
%
\author{Ashkan Khakzar\textsuperscript{1},
Sabrina Musatian\textsuperscript{1},
Jonas Buchberger\textsuperscript{1},
\\
Icxel Valeriano Quiroz\textsuperscript{1},
Nikolaus Pinger\textsuperscript{1},
Soroosh Baselizadeh\textsuperscript{1},
\\
Seong Tae Kim\textsuperscript{2,${\ast}$,}\thanks{denotes corresponding author (st.kim@khu.ac.kr), $^{\ast}$ S. T. Kim and N. Navab shared senior authorship.},
Nassir Navab\textsuperscript{1,3,${\ast}$}}
%
%
%
\authorrunning{ }
%
\institute{\textsuperscript{1}Technical University of Munich \\
\textsuperscript{2}Kyung Hee University\\
\textsuperscript{3}Johns Hopkins University}
\maketitle              
\begin{abstract}
Convolutional neural networks are showing promise in the automatic diagnosis of thoracic pathologies on chest x-rays.  
Their black-box nature has sparked many recent works to explain the prediction via input feature attribution methods (aka saliency methods). However, input feature attribution methods merely identify the importance of input regions for the prediction and lack semantic interpretation of model behavior.  
In this work, we first identify the semantics associated with internal units (feature maps) of the network.  
We proceed to investigate the following questions;  
Does a regression model that is only trained with COVID-19 severity scores implicitly learn visual patterns associated with thoracic pathologies?  
Does a network that is trained on weakly labeled data (e.g. healthy, unhealthy) implicitly learn pathologies?   
Moreover, we investigate the effect of pretraining and data imbalance on the interpretability of learned features. 
In addition to the analysis, we propose semantic attribution to semantically explain each prediction. 
We present our findings using publicly available chest pathologies (CheXpert~\cite{irvin2019chexpert}, NIH ChestX-ray8~\cite{wang2017chestx}) and COVID-19 datasets (BrixIA~\cite{signoroni2020end}, and COVID-19 chest X-ray segmentation dataset~\cite{blockchain}). The Code\footnote{\url{https://github.com/CAMP-eXplain-AI/CheXplain-Dissection}} is publicly available.

\keywords{Interpretability  \and COVID-19 \and Chest X-rays.}
\end{abstract}

\section{Introduction}
Convolutional neural networks (CNN) have demonstrated outstanding performance in automatic diagnosis on Chest X-rays~\cite{wang2017chestx,li2018thoracic,johnson2019mimic,irvin2019chexpert}. There are reports of CNNs outperforming radiologists in chest x-ray pathology classification~\cite{rajpurkar2017chexnet}. These diagnostic models can aid the clinicians and expedite the diagnosis resulting in more patients receiving the care they need. Such models can be especially beneficial in pandemic circumstances as the shortage of expert clinicians becomes an issue~\cite{signoroni2020end,oh2020deep,punn2020automated}. Despite their performance, neural networks’ lack of interpretability undermines their reliability. It is essential to understand the basis of the network predictions, and the networks' learned features to establish trust in the clinical domain. Therefore there have been efforts in explicitly making the models more interpretable during training~\cite{taghanaki2019infomask,khakzar2019learning}, or interpreting neural network models after they are trained~\cite{wang2017chestx,rajpurkar2017chexnet,khakzar2021explaining,karim2020deepcovidexplainer}. In this work, we investigate the latter case in order to see if these performant models trained without an infusion of interpretability, are learning human-interpretable concepts.

For post-hoc explanation of chest X-ray models, many works opt for feature attribution methodologies ~\cite{simonyan2013deep,sundararajan2017axiomatic,bach2015pixel,selvaraju2017grad,khakzar2021neural,wang2017chestx}. These works use feature attribution methods, such as Class Activation Maps (CAM)~\cite{wang2017chestx,selvaraju2017grad} to reveal which input regions are contributing to the output prediction.
Albeit being insightful, the aforementioned methodology lacks semantic interpretation of the models and their predictions. Network Dissection~\cite{bau2017network} is a methodology for identifying the corresponding concept of internal units (feature maps) of the network.

\subsubsection{Contributions -}
In this work, we first use Network Dissection~\cite{bau2017network} to quantify the interpretability of chest X-ray classification models. Then we proceed to investigate the following; 
Does a neural network regression model that is only trained on COVID-19 severity scores (on BrixIA) implicitly learn visual patterns associated with thoracic pathologies? We also study the effect of pretraining on CheXpert and ImageNet datasets, and the effect of considering data imbalance on the semantics of internal units.
Does a network trained on a weakly labeled dataset (healthy/unhealthy labels) implicitly learn distinct pathologies? We combine NIH ChestX-ray8, CheXpert, and BrixIA datasets to generate a massive but weakly labeled ('healthy','unhealthy') dataset. In this case, we study the effect of considering data imbalance on the semantics of internal units.
Moreover, for both cases, we observe the formation of semantic units during training. In all experiments, we use bounding boxes in the NIH ChestX-ray8~\cite{wang2017chestx}, and segmentation masks in COVID-19 chest X-ray segmentation dataset~\cite{blockchain} for identifying the semantics of units.
In addition to the analysis, we propose semantic attribution by combining feature attribution and network dissection to semantically explain the prediction for each chest x-ray.
 
\subsubsection{Related Work -}
 
\textbf{Interpreting internal units:} There are two principal categories of methods for this purpose; Methods that generate images that maximally activate neurons/units~\cite{nguyen2016synthesizing,olah2017feature,simonyan2013deep}, and methods that search over the dataset to find which images (also image regions) activates neurons/units~\cite{zeiler2014visualizing,bau2017network}. Methods in the first category are prone to subjective interpretation as the generated images are ambiguous.
\textbf{Network Dissection}~\cite{bau2017network} is a prominent method of the second category, that does quantitative analysis of the semantics of units. Though we use the same method (albeit on the different domain of chest x-rays rather than natural images), our experiments and insights differ. Effect of pretraining, imbalanced datasets, studying regression models and trained models on weakly-labeled datasets are exclusive to our work. We also propose semantic attribution. 
\textbf{DeepMiner}~\cite{wu2018deepminer} is a methodology inspired by network dissection and applied to mammograms. The methodology differs from Network Dissection and our work. In DeepMiner, instead of automatic annotation, the most important units are annotated by experts. DeepMiner thus differs in methodology, discusses a different domain (mammograms), and does not address our research questions.

\section{Methodology}
\subsection{Setup: Chest X-ray Models:} 

\textbf{Classification model: } Each chest X-ray can contain multiple pathologies. Therefore  we model the problem as a multi-label classification problem. For $C$ pathologies, the network function is defined as $\mathbf{f}_{\Theta}(\mathbf{x}):\mathbb{R}^{H \times W} \rightarrow \mathbb{R}^{C}$. The predicted probability for each category is $\hat{\mathbf{y}} = sigmoid(\mathbf{f}_{\Theta}(\mathbf{x}))$. We use Binary Cross Entropy (BCE) loss on each output. Thus the loss is defined as:

\begin{equation}
\mathcal{L}_{BCE} = (\hat{\mathbf{y}}, \mathbf{y}) = - \sum_{c} \beta \mathbf{y_{c}}\, \log (\hat{\mathbf{y_{c}}}) + (1-\mathbf{y_{c}})\, \log (1-\hat{\mathbf{y_{c}}}) 
\label{eq:loss}
\end{equation}
where $\beta$ is a weighting factor to balance the positive labels, and defined as the ratio of the number of negative labels to the number of positive labels in a batch. 
\\ 
\\
\textbf{Regression model:} We consider two regression modelings on BrixIA dataset.
\\
1) For each input $j$ the label is one global sevirity score $\mathbf{y}^{(j)} \in \{0,...,18\}$. For this case the neural network function is $\mathbf{f}_{\Theta}(\mathbf{x}):\mathbb{R}^{H\times W} \rightarrow \mathbb{R}$. We use a weighted Mean Square Error (MSE) loss for a batch of size N:
\begin{equation}
    \mathcal{L}_{MSE} = \frac{1}{N} \sum \beta(\mathbf{\mathbf{f}_{\Theta}(\mathbf{x})^{(n)}}-\mathbf{y^{(n)}})^{2}
\end{equation}
\\
2) In BrixIA dataset for each X-ray, the lung is divided into 6 regions, and a severity score $\in \{0,1,2,3 \}$ is assigned to each region. Thus the network is defined by $\mathbf{f}_{\Theta}(\mathbf{x}):\mathbb{R}^{H\times W} \rightarrow \mathbb{R}^{6}$. Similar to \cite{signoroni2020end} we use a mixed regression/classification loss by using Sparse Categorical Cross Entropy (SCCE) and diffentiable Mean Absolue Error ($MAE^{d}$), and for each sample and its corresponding network output $\hat{\mathbf{y}} = [\hat{\mathbf{y}}_{c}]^{6}$ it is defined as $ \mathcal{L}_{SCCE} + \mathcal{L}_{MAE^{d}} $:

\begin{equation}
    \mathcal{L}_{SCCE}=-\frac{1}{C}\sum_{c}^{}\mathbf{y}_{c}log(\hat{\mathbf{y}}_{c})
\end{equation}
\begin{equation}
     \mathcal{L}_{MAE^{d}}=\frac{1}{C}\left \| \mathbf{y}-\sum_{c}^{}\frac{e^{\hat{\mathbf{y}_{c}}}}{\sum_{c}^{} e^{\hat{\mathbf{y}_{c}}}}c \right \|
\end{equation}

\subsection{Background: Network Dissection~\cite{bau2017network}}
\begin{figure}[t]
\includegraphics[width=\textwidth]{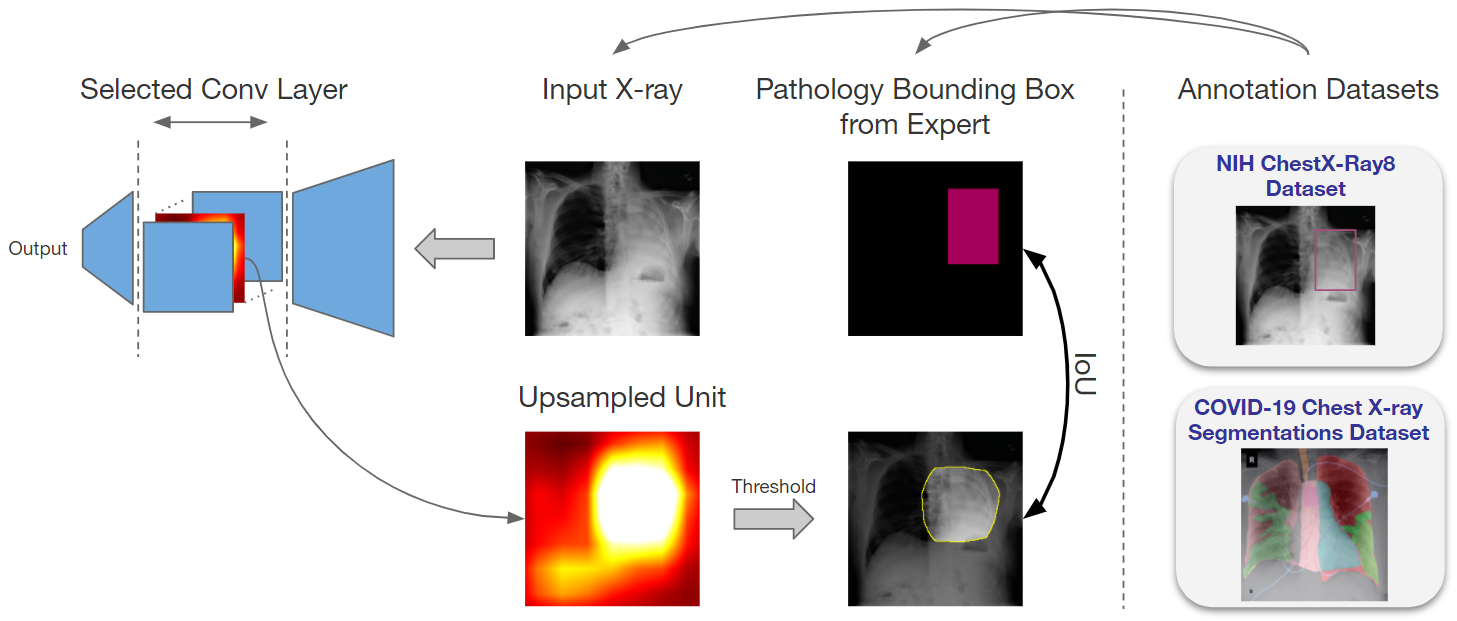}
\caption{\textbf{Network Dissection on chest X-rays:}  
To identify the corresponding concept of a unit in the network, the unit's activation for each input is compared with the ground truth mask/bounding-box of the concept in that input. For one unit the procedure is repeated for all inputs within the annotation datasets and average IoU with concepts in the dataset measures whether the unit is detecting a concept.
} 
\label{fig:method}
\end{figure}

\begin{figure}[t]
\includegraphics[width=\textwidth]{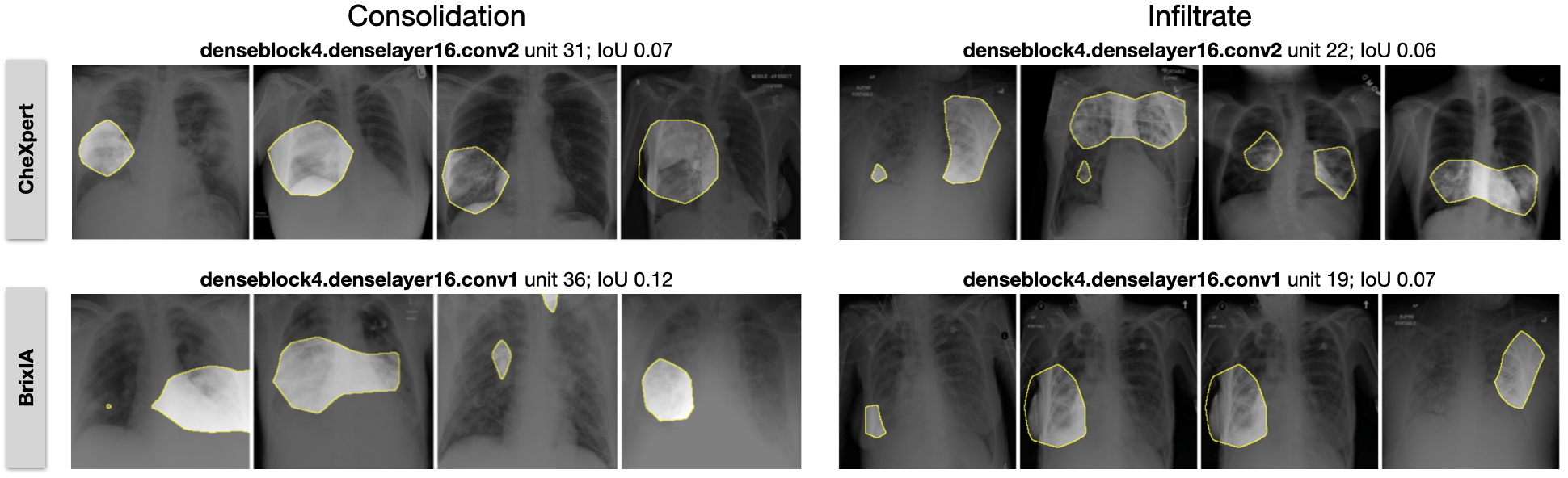}
\caption{\textbf{Individual concept detectors:} For CheXpert classification model (top) and BrixIA regression model (bottom) we visualize a Consolidation detector unit and an Infiltration detector unit for 4 different X-rays. The displayed IoU is not the IoU between the ground truth sample and the detector, it is the average from Eq. \ref{eq:iou}.}
\label{fig:individual_detector}
\end{figure}

Network Dissection annotates individual units (feature maps) of neural networks with semantic concepts. We refer to a feature map that is associated with an individual concept (e.g. consolidation) as an individual concept detector / a semantic unit. The method requires a dataset(s) where concepts are annotated with bounding box or segmentation masks. For each unit under investigation, the unit's feature map is computed for every image and is compared against the ground truth annotations (bounding boxes or segmentation masks) of the concepts in that image. The method compares the unit's feature map and the annotation in terms of their intersection over union. See Fig. \ref{fig:method} for a schematic overview. The activation map is first transformed in to a binary segmentation mask $S_{k}(\mathbf{x})$ via thresholding. For each unit $k$ and its feature map $F_{k}(\mathbf{x})$, the threshold $T_{k}$ is chosen such that $P(f_{k} > T_{k})=0.005$ \cite{bau2017network} for every $f_{k}$ given all images in the dataset. As the resolution of ground truth labels $L_{c}$ for each concept $c$ is different from $S_{k}$, bilinear interpolation is first applied to generate $S'_{k}(\mathbf{x})$. The binary segmentation mask is then derived by $M_{k}(\mathbf{x}) \doteq S'_{k}(\mathbf{x})\geq T_{k}$. For each unit $k$ and concept $c$ pair, the intersection and the union between mask $M_{k}(\mathbf{x})$ derived from unit $k$  and the label mask $L_{c}(\mathbf{x})$ is computed for all images containing concept $c$, and data-set-wide $IoU_{k,c}$ is defined as:
\begin{equation}
\label{eq:iou}
    IoU_{k,c} = \frac{\sum |M_{k}(\mathbf{x})\cap L_{c}(\mathbf{x})|}{\sum |M_{k}(\mathbf{x}) \cup L_{c}(\mathbf{x})|}
\end{equation}
the sum is carried out over all the images with concept $c$, and $|.|$ denotes the cardinality of this set. The $IoU_{k,c}$ measures whether unit $k$ detects concept $c$. We use a threshold of 0.04 similar to \cite{bau2017network} for considering a unit as a detector. The threshold affects the number of concept detectors within a network, what is of interest is \emph{comparing} the number of concept detectors between models (e.g. comparing trained and initial model). See Fig. \ref{fig:individual_detector} for units of different $IoU_{k,c}$. 

\subsubsection{Network Dissection for Chest X-ray Models:} 
We use the NIH ChestX-ray8~\cite{wang2017chestx} and COVID-19 chest X-ray segmentation~\cite{blockchain} (Covid-CXR) dataset for annotating the chest X-ray models. NIH ChestX-ray8 contains bounding boxes for 8 pathologies, and we consider each pathology as a concept $c$ in the Network Dissection framework. Covid-CXR contains segmentations for pathologies, lung components (e.g. right lung), and apparatus. we consider each one as a concept. The dataset contains 14 distinct concepts in total. For each model that we discuss in the paper, we use both NIH ChestX-ray8 and Covid-CXR datasets for annotating the models' internal units.

\begin{figure}[h]
\centering
\includegraphics[width=\textwidth]{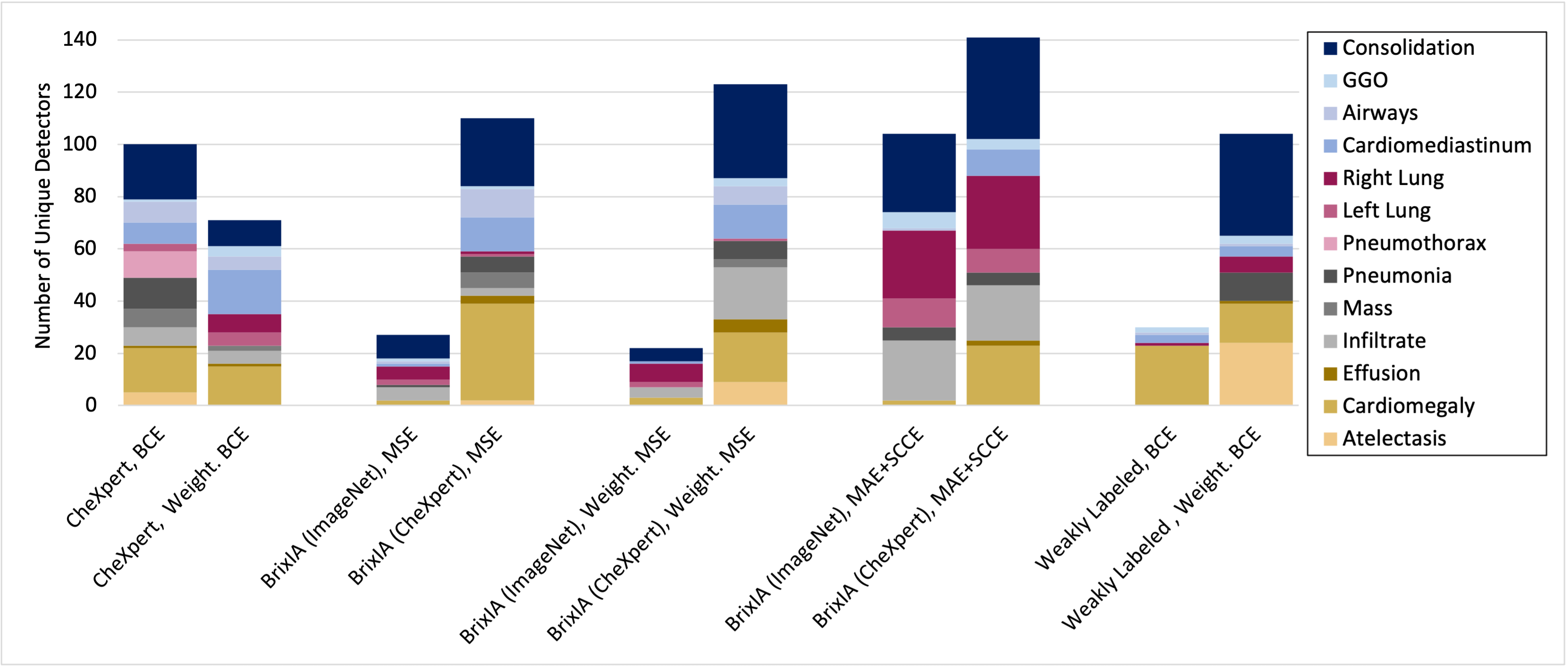}
\caption{\textbf{The number/type of individual concept detectors:} For all models under discussion, the number of concept detectors and the types of the detectors are presented. We can observe that pretraining on CheXpert is increasing the number and the variety of detectors. Moreover, considering data imbalance by using a weighted loss function also increases the number and variety of semantic detectors.} 
\label{fig:quant}
\end{figure}

\subsection{Semantic Attribution}  \label{sec:attribution}
For each prediction $\mathbf{f}_{\Theta}(\mathbf{x})$ on each input $\mathbf{x}$, we compute the importance of internal units $A_{k}^{L}$ for the prediction. $A_{k}^{L}$ denotes $k_{th}$ feature map at layer $L$.  

We follow a method, Integrated Gradients~\cite{sundararajan2017axiomatic}, that approximates the Aumann-Shapley value~\cite{sundararajan2020many}. Shapley value~\cite{shapley1953value} is the unique axiomatic definition of a feature's contribution to a final outcome. We first compute the individual neurons' importance $a_{k}^{i}$ (activation $i$ at channel $k$):

\begin{equation}
    \mathbf{s}^{i}_{k}=\mathbf{a}^{i}_{k}\int_{\alpha = 0}^{1} \frac{\partial \mathbf{f}_{\theta} (\alpha \mathbf{a}^{i}_{k}) }{\partial \mathbf{a}^{i}_{k}} \mathrm{d}\alpha
\end{equation}

\noindent We use $\sum_{i}^{}|\mathbf{s}_{k}^{i}|$ as the contribution of the unit $k$. The absolute value is for considering both positive and negative contributions by taking the magnitude. We then select the top contributing unit(s) and use their annotations (if it exists) for semantic explanation. We visualize $\mathbf{s}^{i}_{k}$ of all neurons of $A_{k}^{L}$ to highlight each semantic unit's most contributing areas to the output.

\begin{figure}[t]
\includegraphics[width=\textwidth]{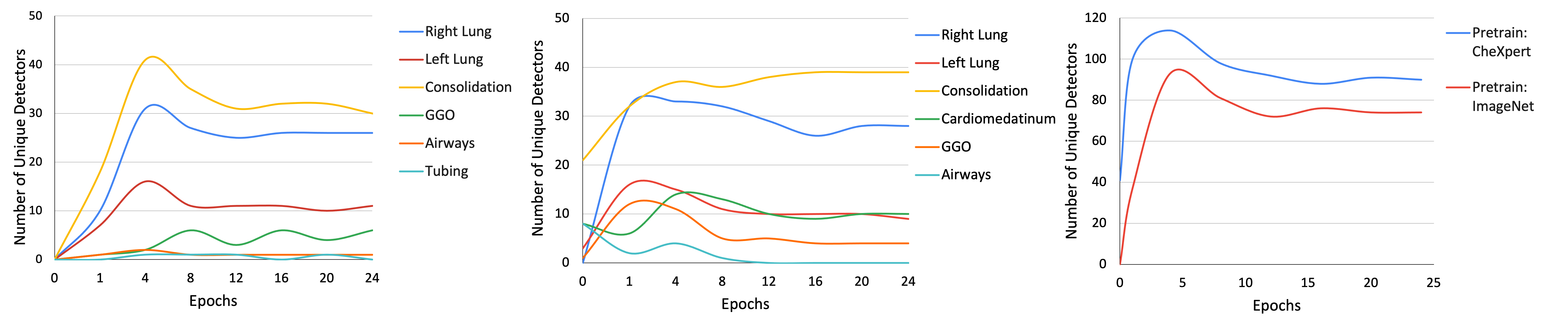}
\caption{\textbf{Evolution of concepts:} Number/Type of semantic detectors in two regression models at different epochs. (left): BrixIA (ImageNet), MAE+SCCE (middle): BrixIA (CheXpert), MAE+SCCE, (right) total number of semantic detectors for both models. We observe that the number of concept detectors is higher for trained models at different epochs compared to the initial model.} 
\label{fig:evolution}
\end{figure}

\section{Results and Discussion}
In all our experiments and for all models in Fig. \ref{fig:quant}, the network is a DenseNet121 \cite{DBLP:journals/corr/HuangLW16a}. All models are trained using Adam \cite{kingma2014adam} optimizer without weight decay. If the loss function is a weighted one, it is indicated by "Weight".
For BrixIA models, the model is trained on a pre-trained model on either ImageNet or CheXpert. The name in parentheses shows this. When the last part of the name shows the loss function. "MAE+SCCE" indicates using the 6 region prediction as in the original BrixIA paper. The weakly labeled models are binary classifiers, trained on the combination of NIH, CheXpert, and BrixIA datasets, deciding if the input is healthy or not. While NIH dataset has a healthy label for inputs, for CheXpert, the input was considered healthy when no pathology existed. For BriXIA, an input was considered healthy when all 6 regions were zero.

\subsection{Semantics of Thoracic Classification Models}
In this section, we analyze the semantics of detectors in classification models trained on CheXpert. Although these models are trained directly on the pathology labels, this does not imply that the individual detectors within the network are detecting features relevant to these pathologies. The models are denoted by CheXpert BCE, and CheXpert Weight. BCE on Fig. \ref{fig:quant}.

\subsubsection{Balanced Training Effect}

We see that the model trained with weighted BCE has a significantly higher number of concept detectors (100 vs 65).  The number of detectors relevant to Pneumonia increases significantly in the weighted model. Pneumonia has a few positive instances in the dataset (0.027\%), using a weighted loss causes the emergence of Pneumonia detectors. Consolidation is also imbalanced (0.066\%), and we observe that the number of detectors is tripled. In addition, the weighted model shows more variety in terms of concepts than the unweighted version. This is particularly significant as we do not observe any improvement in terms of classification AUC, and also the F1 score. But the model interpretability increases with the weighted loss.

\subsection{Semantics of COVID-19 Regression Models}
In this section, we examine MSE and MAE SCCE segmentation losses for models trained on the BrixIA dataset. In addition, we investigate the effect of considering imbalance in the data in the MSE loss models. As the number of images in BrixIA is limited, we also investigate the effect of pretraining on the CheXpert dataset. The first observation in Fig. \ref{fig:quant} is that training the model on 6 region regression loss (SCCE), results in a higher number of semantic detectors compared to global regression loss. The model has stronger supervision in the former case, and we observe that it also becomes more interpretable.

\subsubsection{Pretraining Effect}
For all training losses, we observe in Fig. \ref{fig:quant} that pretraining the model on the CheXpert dataset is significantly increasing the number of individual concept detectors. This increase in the number of detectors is more pronounced in models trained with the global regression score labels. This implies that with weaker supervision with one global regression score, the model struggles in learning individual concepts, and pretraining becomes necessary.


\subsubsection{Balanced Training Effect}
We observe (Fig. \ref{fig:quant}) that using weighted MSE loss on regression models pre-trained on CheXpert datasets increases the number of unique concept detectors, specifically an increase in the number of detectors associated with Consolidation. For models that are not pretrained on CheXpert, the model struggles in both weighted and unweighted scenarios to learn concepts.

\subsection{Semantics of Models Trained on Weakly Labeled Datasets}
In this section we discuss models trained on a dataset of healthy / unhealthy labels. The dataset is comprised of CheXpert, NIH ChestX-ray8 and BrixIA.

\subsubsection{Balanced Training Effect}
We observe in Fig. \ref{fig:quant} that the model that does not use a weighted loss, learns few semantic concepts. Moreover, the majority of individual detectors are related to Cardiomegaly and Cardiomediastinum, and the rest of the concepts are not learned. This implies that Cardiomegaly is an easy signal for the model to pick up during training. However, when we train the model with a weighted loss, we observe that the diversity of the semantic detectors and their numbers increases significantly. Specifically, there were not Consolidation, Pneumonia, Atelectasis detectors in the former case, but they emerge in high numbers in the weighted case. We also observe that the concept detectors indeed emerge in models trained on healthy/unhealthy labels, although it is relatively less than full supervision.

\subsection{Evolution of Semantics}
In this section, we observe the number/type of individual unit detectors for different epochs. We analyze the BrixIA regression models trained with MAE+SCCE. 
We observe (Fig. \ref{fig:evolution}) an increase in the number of important concepts for COVID-19 during training (e.g. Consolidation and Ground Glass Opacity) and a decrease in less important concepts (e.g. left lung). The relative increase in the number of concepts of trained model vs. initial model points to the fact that the models are indeed learning patterns associated with pathologies.

\begin{figure}[t]
\includegraphics[width=\textwidth]{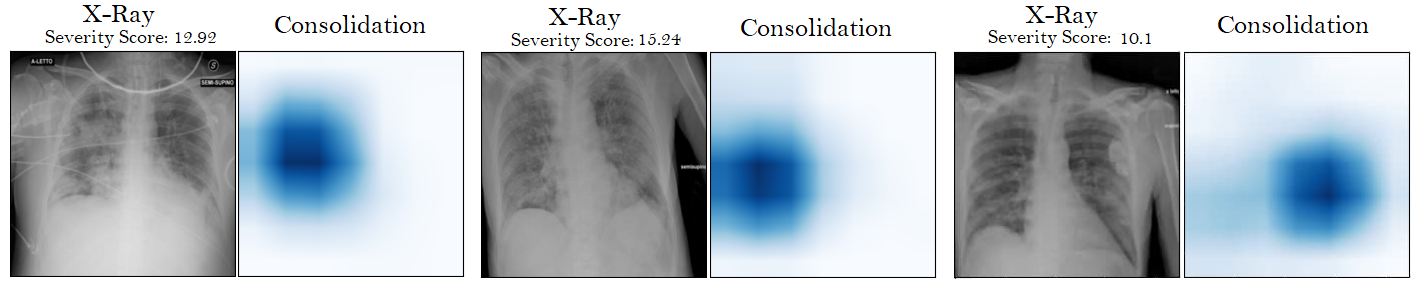}
\caption{\textbf{Semantic attribution:} X-rays with severe Covid condition from BrixIA. All have a unit with corresponding semantic annotation of Consolidation as their top contributing unit to the severity score. The contribution of corresponding unit is visualized.
}
\label{fig:attribution}
\end{figure}
\subsection{Semantic Attribution}
We observe that Consolidation is the most related concept to severe Covid scores as in Fig. \ref{fig:attribution} on BrixIA dataset. The results in Fig. \ref{fig:attribution} are for the last conv layer of the model described in Fig. \ref{fig:quant} as BrixIA (CheXpert), Weight, MSE. To visualize, the final integrated gradients map of unit $k$ is divided by the maximum $\mathbf{s}_{k}^{i}$ of its neurons to normalize and then upsampled to the input image size. Note the methodology here is different from using crude activation maps since we aim to discover the most \textit{contributing} feature maps to the prediction. Highly activated maps do not necessarily correspond to contribution. Hence, we employ an axiomatic approach as explained in Sec. \ref{sec:attribution}. Although this is a first step in this direction, more annotations can boost the recognition of other important concepts, and more semantic units for each prediction can be visualized.

\section{Conclusion}
In this work, we analyze the semantics of internal units of classification and regression models on chest X-rays. We observe that pretraining and considering data imbalance affect the number of semantic detectors. We observe how training on severity score regression labels results in the emergence of pathology-related detectors within the networks. We observe the same phenomenon when the model is trained on weakly labeled datasets. We also propose a semantic attribution method to semantically explain individual predictions.

\subsubsection{Acknowledgement}
This work is partially funded by the \textbf{Munich Center for Machine Learning (MCML)} and the \textbf{Bavarian Research Foundation} grant AZ-1429-20C. The computational resources for the study are provided by the \textbf{Amazon Web Services Diagnostic Development Initiative}.
S.T. Kim is supported by the \textbf{Korean MSIT}, under the National Program for Excellence in SW (2017-0-00093), supervised by the IITP.

\bibliographystyle{splncs04}
\bibliography{references}
\end{document}